\documentstyle[amsfonts,referee]{mn}

\def\ST{{$({\cal M},\gten)$\ }}			% mathematical short for spacetime
\def\AS{{$(\Sigma_t,\tgamma)$\ }}		% mathematical short for absolute 3-space

\def\rangles{{\rangle^{\rm{s}}}}
\def\ranglea{{\rangle^{\rm{a}}}}
\def\upara{{\uten}}
\def\uperp{{\hten}}
\def\vpara{{\tgv}}
\def\vperp{{\ten h}}

\def\mDelta{{\mit\Delta}}		% Kerr metric function
\def\mPi{{\mit\Pi}}			% viscous bulk pressure
\def\mSigma{{\mit\Sigma}}		% scattering cross section
\def\mTheta{{\mit\Theta}}		% expansion of fluid trajectories

\ifCUPmtlplainloaded \else
  \DeclareSymbolFont{UPM}{U}{eur}{m}{n}
  \SetSymbolFont{UPM}{bold}{U}{eur}{b}{n}
  \DeclareMathSymbol{\ubeta}{0}{UPM}{"0C}
  \DeclareMathSymbol{\ueta}{0}{UPM}{"11}
  \DeclareMathSymbol{\ugamma}{0}{UPM}{"0D}
  \DeclareMathSymbol{\uomega}{0}{UPM}{"21}
  \DeclareMathSymbol{\upartial}{0}{UPM}{"40}
  \DeclareMathSymbol{\upi}{0}{UPM}{"19}
  \DeclareMathSymbol{\usigma}{0}{UPM}{"1B}
  \DeclareMathSymbol{\utheta}{0}{UPM}{"12}
  \DeclareMathSymbol{\uvartheta}{0}{UPM}{"23}
\fi
 
\def\aten{{\rm a}}		% acceleration vector
\def\Bten{{\rm B}}		% arbitrary vector/tensor field
\def\Cten{{\rm C}}		% arbitrary vector/tensor field
\def\Dten{{\rm D}}		% covariant spatial derivative operator (vector type)
\def\dten{{\rm d}}		% exterior derivative operator (vector type)
\def\eten{{\rm e}}		% basis vector
\def\Gten{{\rm G}}		% Einstein tensor
\def\gten{{\rm g}}		% metric tensor
		% ...
\def\Kten{{\rm K}}		% extrinsic curvature tensor (=second fundamental form)
\def\kten{{\rm k}}		% axial Killing vector
\def\hten{{\rm h}}		% projection tensor into LRF of fluid
\def\mten{{\rm m}}		% time Killing vector
\def\nten{{\rm n}}		% particle flux vector
\def\qten{{\rm q}}		% heat flux vector
\def\Rten{{\rm R}}		% dissipative contribution to entropy flux vector
\def\sten{{\rm s}}		% entropy flux vector
\def\Tten{{\rm T}}		% stress energy tensor
\def\uten{{\rm u}}		% velocity vector
\def\Yten{{\rm Y}}		% constraint help vector
\def\Zten{{\rm Z}}		% constraint help tensor

\def\gammaten{\ugamma}		% projector into FIDO space
\def\omegaten{\uomega}		% vorticity tensor
\def\piten{\upi}		% anisotropic viscous stress tensor
\def\sigmaten{\usigma}		% shear tensor
\def\thetaten{\utheta}		% basis 1-form

\def\SIPi{{\tilde\mPi}}			% SI bulk pressure
\def\SIqten{{\tilde\qten}}      	% SI heat flux vector
\def\SIpiten{{\tilde\piten}}		% anisotropic SI viscous stress tensor

\def\GTheta{{\underline{\mTheta}}}      % G (=geodesic) expansion
\def\Gaten{{\underline{\aten}}}         % G (=geodesic) acceleration (=0)
\def\Gsigmaten{{\underline{\sigmaten}}} % G (=geodesic) shear

\def\GSIPi{{\underline{\SIPi}}}		% GSI bulk pressure
\def\GSIqten{{\underline{\SIqten}}}	% GSI heat flux vector
\def\GSIpiten{{\underline{\SIpiten}}}	% anisotropic GSIviscous stress tensor

\def\GPi{{\underline{\mPi}}}
\def\Gqten{{\underline{\qten}}}
\def\Gpiten{{\underline{\piten}}}

\def\haten{{\hat{\rm a}}}			% FIDO 4-acceleration
\def\hnten{{\hat{\rm n}}}			% FIDO 4-velocity
\def\homegaten{{\hat{\uomega}}}			% FIDO 4-vorticity
\def\hsigmaten{{\hat{\usigma}}}			% FIDO 4-shear
\def\hTheta{{\hat\mTheta}}                      % FIDO 4-expansion

\def\Ddot{{\dot{D}}}				% rate-of-change w.r.t. LRF
\def\Dbul{{\dot{\cal D}}}			% rate-of-change w.r.t. FIDO

\def\beq{\begin{equation}}
\def\eeq{\end{equation}}
\def\bea{\begin{eqnarray}}
\def\eea{\end{eqnarray}}
\def\beas{\begin{eqnarray*}}
\def\eeas{\end{eqnarray*}}
\def\bit{\begin{itemize}}
\def\eit{\end{itemize}}

\def\tenG #1{\mbox{\boldmath$#1$}}	% yields bold italics, also for lower-case greek
\def\ten #1{\bf#1}			% yields bold upright, except for loweri-case greek

\def\te{{\ten e}}		% spatial basis vector				

\def\ta{{\ten a}}		% sational acceleration of FIFO
\def\tD{{\ten D}}		% arbitrary spatial tensor field
\def\tF{{\ten F}}		% spatial help function vector field
\def\tG{{\ten G}}               % gravitational acceleration measured by FIDO 
\def\tH{{\ten H}}		% grovitomagnetic tensor field
\def\tJ{{\ten J}}		% grovitomagnetic vector field
\def\tK{{\ten K}}		% spatial extrinsic curvature tensor field
\def\tm{{\ten m}}		% spatial axial Killing vector
\def\tv{{\ten v}}		% spatial fluid velocity field devided by Lorentz factor
\def\tS{{\ten S}}		% spatial current vector field
\def\tT{{\ten T}}		% spatial stress-tensor field
\def\tgv{{\ten u}}		% spatial fluid velocity field
\def\tv{{\ten v}}		% spatial fluid velocity field devided by Lorentz factor

\def\Whelp{{W}}			% scalar help function
\def\tWhelp{{\ten\Whelp}}	% spatial help function vector field

\def\bbbr{{\rm I\!R}} 		% real number symbol taken from A&A l-aa.sty

\def\tbeta{{\tenG\ubeta}}			% 3-shift 
\def\tgamma{{\tenG\ugamma}}			% 3-metric
\def\tnabla{{\tenG\nabla}}			% 3-gradient
\def\ttheta{{\tenG\uvartheta}}			% 3-basis one-form
\def\tomega{{\tenG\uomega}}			% 3-vorticity
\def\tsigma{{\tenG\usigma}}			% 3-shear

\def\fraca{{\frac{1}{\alpha}}}
\def\fractau#1{{\frac{1}{\tau_{#1}}}}
\def\gfraca{{\frac{\gam}{\alpha}}}
\def\DConLie{{\fraca\left(\part-\Liebeta\right)}}
\def\gDConLie{{\gfraca\left(\part-\Liebeta\right)}}
\def\DConbeta{{\fraca\left(\fpart-\tbeta\cdot\tnabla\right)}}
\def\gDConbeta{{\gfraca\left(\fpart-\tbeta\cdot\tnabla\right)}}

\def\homega{{\hat\uomega}}
\def\hsigma{{\hat\usigma}}			
\def\tha{{\ten{\hat a}}}			% FIDO 3-acceleration
\def\thomega{{\tenG{\homega}}}			% FIDO 3-vorticity
\def\thsigma{{\tenG{\hsigma}}} 			% FIDO 3-shear

\def\lie{{\cal L}}
\def\Liebeta{{{\lie}_\tbeta}}

\def\gv{u}

\def\tgv{{\ten\gv}}

\def\EX{\mathop{E}}		% with these declarations (see Schwarz, TeX-Buch, p. 133ff)
\def\SX{\mathop{S}}		% subscripts/superscripts appear below/above the symbol
\def\TX{\mathop{T}}
\def\tSX{\mathop{\tS}}	
\def\tTX{\mathop{\tT}}	

\def\TE{\EX_\Tten}
\def\sigmaE{\EX_\sigmaten}
\def\uE{\EX_\uten}
\def\aE{\EX_\aten}
\def\ThetaE{\EX_\mTheta}

\def\AE{\EX_A}
\def\BE{\EX_\Bten}
\def\CE{\EX_\Cten}

\def\PiE{\EX_\mPi}

\def\SIPiE{\EX_\SIPi}
\def\GPiE{\EX_\GPi}
\def\GSIPiE{\EX_\GSIPi}

\def\qE{\EX_\qten}

\def\SIqE{\EX_\SIqten}
\def\GqE{\EX_\Gqten}
\def\GSIqE{\EX_\GSIqten}

\def\piE{\EX_\piten}

\def\SIpiE{\EX_\SIpiten}
\def\GpiE{\EX_\Gpiten}
\def\GSIpiE{\EX_\GSIpiten}

\def\qtaE{\EX_{\qten\cdot\aten}}
\def\pitaE{\EX_{\piten\cdot\aten}}

\def\BS{\mathord{\SX_\Bten}}
\def\CS{\mathord{\SX_\Cten}}

\def\TtS{\tSX_\Tten}
\def\sigmatS{\tSX_\sigmaten}
\def\utS{\tSX_\uten}
\def\atS{\tSX_\aten}

\def\BtS{\tSX_\Bten}
\def\CtS{\tSX_\Cten}

\def\qtS{\tSX_\qten}

\def\SIqtS{\tSX_\SIqten}
\def\GqtS{\tSX_\Gqten}
\def\GSIqtS{\tSX_\GSIqten}

\def\pitS{\tSX_\piten}

\def\SIpitS{\tSX_\SIpiten}
\def\GpitS{\tSX_\Gpiten}
\def\GSIpitS{\tSX_\GSIpiten}

\def\pitatS{\tSX_{\piten\cdot\aten}}

\def\CT{\mathord{\TX_\Cten}}

\def\TtT{\tTX_\Tten}
\def\sigmatT{\tTX_\sigmaten}

\def\CtT{\tTX_\Cten}

\def\pitT{\tTX_\piten}

\def\SIpitT{\tTX_\SIpiten}

\def\GpitT{\tTX_\Gpiten}
\def\GSIpitT{\tTX_\GSIpiten}

\def\gam{{\gamma}}

\def\tr{\,\rm{tr}}				% the trace is a scalar
\def\exp #1{{\rm e}^{#1}}			% exponential function, in LateX just \exp=\rm exp
\def\tilomega{{\tilde\omega}}
\def\cad2{{c^2_{\rm s}}}
\def\cvis2{{c^2_{\rm v}}}

\def\partial{{\upartial}}
\def\part{{\partial_t}}

\def\fpart{{\frac{\partial}{\partial t}}}

\def\sks{\;\;,\;\;}

\begin{document}

\title{3+1 formulation of non-ideal hydrodynamics}

\author[J. Peitz and S. Appl]
{Jochen Peitz$^1$\thanks{e-mail: {\tt jpeitz@lsw.uni-heidelberg.de}}
and 
Stefan Appl$^2$$^,$$^3$\thanks{e-mail: {\tt appl@gaia.iwr.uni-heidelberg.de}}
\\
$^1$Landessternwarte K\"onigstuhl, 69117 Heidelberg, Germany\\
$^2$Observatoire Astronomique, Universit\'{e} Louis Pasteur, 
11, rue de l'Universit\'{e}, 67000 Strasbourg, France\\
$^3$Institut f\"ur Angewandte Mathematik, 
Universit\"at Heidelberg, Im Neuenheimer Feld 294, 69120 Heidelberg, Germany}
\date{Accepted for publication in the MNRAS}
\maketitle

%Accepted for publication in the MNRAS
\begin{abstract}
The equations governing dissipative relativistic hydrodynamics
are formulated within the 3+1 approach for arbitrary spacetimes.
Dissipation is accounted for by applying the theory of extended causal
thermodynamics (Israel-Stewart theory).
This description eliminates the causality violating infinite signal speeds
present in the conventional Navier-Stokes equation.
As an example we treat the astrophysically relevant case of stationary
and axisymmetric spacetimes, including the Kerr metric. The equations
take a simpler form whenever the inertia due to the dissipative
contribution can be neglected.
%The set of equations is specified to the astrophysically relevant case of
%stationary axisymmetric vacuum spacetimes, including the Kerr metric.
%The equations are also given in a simplified form suitable for
%applications where inertia due to the dissipative contributions is negligible.
\end{abstract}

%
%%%%%%%%%%%%%%%%%%%%%%%%%%%%%%%%%%%%%%%%%%%%%%%%%%%%%%%%%%%%%%%%%%%%%%%%%%%%%%%%%%%%%%%
% Introduction
%%%%%%%%%%%%%%%%%%%%%%%%%%%%%%%%%%%%%%%%%%%%%%%%%%%%%%%%%%%%%%%%%%%%%%%%%%%%%%%%%%%%%%%
%
\section{Introduction}
\label{intro}
The motion of dissipative fluids in strong gravitational fields is of
considerable interest in various fields of astrophysics and cosmology.
Examples include accretion discs around compact objects, rotating relativistic fluid
configurations such as supermassive stars, neutron stars or strange (boson) stars, 
the collapse of stellar objects and the merging of compact objects.
Examples in cosmology cover inflationary cosmological scenarios
and the evolution of density fluctuations.

The non-stationary modeling of relativistic matter
is most conveniently performed within the 3+1 formalism,
where the equations of motion for gravitational- and matter fields are decomposed 
w.r.t. a congruence of fiducial observers (FIDOs), 
allowing to express time derivatives on a per-unit-universal-time basis. 
The 3+1 representation of the equations for ideal, non-dissipative (relativistic) hydrodynamics
as well as Maxwell's equations for the electric and magnetic fields and Einstein's
equations for the gravitational fields have been discussed by many authors,
both in a cosmological context (e.g. Durrer \& Straumann 1988) and in the case
of black hole spacetimes (e.g. Thorne \& Macdonald 1982).

Modeling dissipative processes requires non-equilibrium or irreversible thermodynamics.
Standard (or classical), irreversible thermodynamics (in the following referred to as
{\it standard thermodynamics}) 
was first extended from Newtonian to relativistic fluids by Eckart \shortcite{Eckart40}. 
However, the Eckart theory, and a variation thereof by Landau \& Lifshitz 
\shortcite{LandauLifshitz59} shares with its Newtonian counterpart serious problems.
Notably that dissipative fluctuations propagate at an infinite speed.
In addition, generic short wavelength secular instabilities driven by dissipative processes
exist \cite{LindblomHiscock83} and finally, no well-posed initial value problem exists
for rotating fluid configurations.

At the origin of these problems in standard thermodynamics is the description
of non-equilibrium via the local equilibrium states alone, i.e.
it is assumed that local thermodynamic equilibrium is established on an infinitely
short time-scale (see e.g. Jou, Casas-V\'{a}zquez \& Lebon 1997 for an introduction).
In extended theories of irreversible thermodynamics the set of thermodynamic variables 
is extended to include the dissipative variables.
This restores causality and stability under a wide range of conditions
\cite{HiscockLindblom83}.
A non-relativistic extended theory was proposed by M\"{u}ller \shortcite{M"uller67},
and was then generalized to the relativistic case by Israel \shortcite{Israel76}
and Stewart \shortcite{Stewart77}.
The extended theory is commonly referred to as {\it causal thermodynamics},
second-order thermodynamics or transient thermodynamics.

The problem of non-causality has recently received attention in the context of 
transonic accretion discs. 
Two different approaches have been proposed to overcome this difficulty.
For steady flow Narayan \shortcite{Narayan92} has established causality by 
calculating the coefficient of kinematic viscosity within an extended version 
of flux limited diffusion theory \cite{LevermorePomraning81}, 
assuming a particular steady state phase-space distribution function for 
the turbulent fluid elements in the disc.
The influence of this modified viscosity coefficient was studied in stationary accretion discs
by Popham \& Narayan \shortcite{PophamNarayan92} and Syer \& Narayan \shortcite{SyerNarayan93}.
A relativitic generalization of the modified viscosity has been proposed,
and used in models of stationary relativistic accretion discs 
(Peitz \& Appl \shortcite{PeitzAppl97}).
A different approach by Papaloizou \& Szuskiewicz \shortcite{PapaloizouSzuskiewicz94},
which is related to a causal description for the thermodynamics, was used by
Kley \& Papaloizou \shortcite{KleyPapaloizou97} in time-dependent models for 
accretion disc boundary layers.
Gammie \& Popham \shortcite{GammiePopham97} have recently considered 
a similar extension to stationary relativistic accretion discs.
In cosmology the theory of causal thermodynamics is currently attracting growing interest
predominantly in the contexts of re-heating processes after inflation 
\cite{Zimdahletal97}
and in linear perturbation theory for the evolution of density fluctuations
\cite{MaartensTriginer97}.

This paper provides a complete set of equations for dissipative fluid mechanics 
in their 3+1 representation, using a causal description of thermodynamics.
In Section \ref{sec2} the basic elements of both standard and causal dissipative hydrodynamics
are reviewed in their spacetime description, and their 3+1 representation
is derived in Section \ref{sec3}.
As a particular application of astrophysical interest we specify the system of equations
given in Section \ref{sec3} to the case of a stationary, axisymmetric back ground 
in Section \ref{sec4}. 
\section{Dissipative relativistic hydrodynamics}
\label{sec2}
The equations of ideal relativistic fluid mechanics and the equations for
dissipative relativistic fluid mechanics in both the standard irreversible
and the extended causal thermodynamics description are reviewed. 
For a detailed discussion see Israel \& Stewart \shortcite{IsraelStewart79},
Hiscock \& Lindblom \shortcite{HiscockLindblom83} or a recent treatment by
Maartens \shortcite{Maartens97}.
\subsection{Notation}
\label{sec2_1}
We use geometrized units such that $c=1=G$\/.
Tensor fields defined on spacetime \ST with metric $\gten$\/ of signature $(-,+,+,+)$\/
appear in roman (e.g. $\uten,\Tten$\/), while scalar functions are in italic.
The velocity $\uten$\/ of the fluid is normalized to $\uten\cdot\uten=-1$\/. 
The tensor $\hten=\gten+\uten\otimes\uten$\/ projects into the 3-space orthogonal to $\uten$\/,
the local rest frame of the fluid (LRF). Total projections (i.e. projection in any free index)
parallel to and orthogonal to $\uten$\/ are denoted by $(\;)_\upara$\/ and $(\;)_\uperp$\/.
If $A$\/, $\Bten$\/ and $\Cten$\/ are a scalar, vector and rank-2 tensor field on \ST,
respectively, then
\bea
A_\upara		&=&	A_\uperp
			=	A\;,\nonumber\\
\Bten_\upara		&=&	B_\upara
			=	\uten\cdot\Bten\sks
\Bten_\uperp		=	\hten\cdot\Bten\;,\nonumber\\
\Cten_\upara		&=&	C_\upara
			=	\uten\cdot\Cten\cdot\uten\sks
\Cten_\uperp		=	\hten\cdot\Cten\cdot\hten
\label{eqProject}\;.
\eea
%where the position of $\uten,\hten$\/ corresponds to contracted index.
%Obviously $(\;)_\upara$\/ always yields a scalar while $(\;)_\uperp$\/ preserves the rank.
Two covariant differential operators $\Ddot$\/ and $\Dten$\/ are defined
as projections of the affine connection $\nabla$\/ on \ST
into directions parallel and orthogonal to $\uten$\/,
\bea
\Ddot		&\equiv&\nabla_\upara
		=	\uten\cdot\nabla
\label{eqNabu}\;,\\
\Dten		&\equiv&\nabla_\uperp
		=	\hten\cdot\nabla
\label{eqNabh}\;.
\eea
%Formally, $\Dten$\/ is a vector field while $\Ddot$\/ is a scalar field.
%Useful properties of $\Dten$\/ and $\Ddot$\/ are listed in appendix \ref{appdx1}.
For 2-tensor fields $\Cten$\/ we further introduce (anti-) symmetrization operators 
$\langle\;\ranglea$\/ and $\langle\;\rangles$\/ by 
$\Cten=\langle\Cten\ranglea$\/ if $\Cten$\/ is anti-symmetric and
$\Cten=\langle\Cten\rangles$\/ if $\Cten$\/ is symmetric and trace-free.
%Note that $(\;)_\uperp$\/ and $\langle\;\rangle$\/ do commute, 
%$\langle\;\rangle_\uperp=\langle\;_\uperp\rangle$\/, as a consequence of the symmetry of $\hten$\/.
Finally, the irreducible decomposition of $\nabla\uten$\/ yields
\bea
\nabla\uten	&=&	\sigmaten
			+\omegaten
			+1/3\mTheta\hten
			-\aten\otimes\uten
\label{eqDecomp}\;,
\eea
with shear $\sigmaten\equiv\langle\nabla\uten\rangle^{\rm s}_\uperp$\/,
vorticity $\omegaten\equiv\langle\nabla\uten\rangle^{\rm a}_\uperp$\/,
acceleration $\aten\equiv\Ddot\uten$\/ and expansion 
$\mTheta\equiv\Dten\cdot\uten=\nabla\cdot\uten$\/ of the fluid trajectories.
\subsection{Perfect fluids}
\label{sec2_2}
A perfect fluid is described by the velocity $\uten$\/, baryon number density $n$\/,
mass-energy density $\rho$\/, isotropic pressure $p$\/ and specific entropy $s$\/,
which are subject to the conservation laws
\bea
0		&=&	\nabla\cdot\nten
\label{eqntenCons}\;,\\
0		&=&	\nabla\cdot\Tten
\label{eqTtenCons}\;.
\eea
The particle current vector $\nten$\/ and the symmetric stress-energy tensor $\Tten$\/ 
are given by
\bea
\nten		&=&	n\uten
\label{eqInten}\;,\\
\Tten		&=&	\rho\uten\otimes\uten
			+p\hten
\label{eqITten}\;.
\eea
The LRF conservation laws for energy and momentum result from projecting (\ref{eqTtenCons})
parallel and orthogonal to $\uten$\/. With (\ref{eqITten}) one can write
$0=(\nabla\cdot\Tten)_\upara$\/ and $0=(\nabla\cdot\Tten)_\uperp$\/ as
\bea
0               &=&     \Ddot\rho
                        +(\rho+p)\mTheta
\label{eqIEnergy}\;,\\
0               &=&     \Dten p
                        +(\rho+p)\aten
\label{eqIMomentum}\;.
\eea
The metric $\gten$\/ is coupled to stress-energy $\Tten$\/ by Einstein's equations
\bea
\Gten           &=&     8\pi\Tten
\label{eqEinstein}\;,
\eea
where $\Gten$\/ is the Einstein tensor.
% and $\pi$\/ should not be confused with the non-equilibrium anisotropic viscous stress $\piten$\/.

Thermodynamic scalar functions are defined in the LRF.
The entropy flux
\bea
\sten		&=&	s\nten
\label{eqIsten}
\eea
is conserved along flow lines (adiabatic flow),
\bea
0		&=&	\nabla\cdot\sten
\label{eqIstenCons}\;.
\eea
The temperature $T$\/ is defined via the Gibbs equation 
\bea
T\dten s		&=&	\dten\left(\rho/n\right)
				+p\dten\left(1/n\right)
\label{eqGibbs}\;,
\eea
where $\dten$\/ is the exterior derivative on \ST. In general, two thermodynamic scalars
are needed as independent variables, which we choose $n$\/ and $\rho$\/.
A scalar equation of state, e.g. $p=p(n,\rho)$\/, closes the system of equations 
(\ref{eqntenCons}), (\ref{eqIEnergy})-(\ref{eqEinstein}) 
for the dynamical variables $\{n$\/, $\rho$\/, $\uten$\/, $\gten\}$\/. 
\subsection{Dissipative fluids}
\label{sec2_3}
Choosing the particle current for dissipative fluids by (\ref{eqInten}) corresponds to 
selecting an average velocity $\uten$\/ such that the particle flux in the associated rest
frame vanishes. This so-called particle frame or Eckart frame \cite{Eckart40} is the 
natural frame in systems where particle number is conserved
(see Israel \& Stewart 1979 for the alternative energy frame description).
The state of the fluid is assumed close to a fictitious thermodynamic equilibrium state,
characterized by the local thermodynamic equilibrium scalars 
$n_0,\rho_0,p_0,s_0,T_0$\/
and the local equilibrium velocity $\uten_0$\/, which in the Eckart frame can
be chosen such that only the pressure $p$\/ deviates from the local equilibrium pressure
$p_0$\/ by the bulk viscous pressure $\mPi=p-p_0$\/, whereas $n=n_0$\/ and 
$\rho=\rho_0$\/. Dropping subscripts 0 allows then to write the general stress-energy tensor 
for dissipative fluids as
\bea
\Tten		&=&	\rho\uten\otimes\uten
			+(p+\mPi)\hten
			+\qten\otimes\uten
			+\uten\otimes\qten
			+\piten
\label{eqVTten}\;,
\eea
where the heat flux $\qten$\/ relative to the particle frame and the anisotropic 
stress tensor $\piten$\/ are orthogonal to $\uten$\/,
\beq
\qten		=	\qten_\uperp\sks
\piten		=	\langle\piten\rangle^{\rm s}_\uperp
\label{eqVProp}\;.
\eeq
Conservation laws again hold for $\nten$\/ and $\Tten$\/.
However, in irreversible thermodynamics the entropy is no longer conserved.
According to the second law,
the rate of entropy generation must therefore be positive definite,
\bea
\nabla\cdot\sten	&\ge&	0
\label{eqVsLaw}\;,
\eea
implying that $\sten$\/ has a dissipative vector contribution $\Rten$\/ in excess to 
(\ref{eqIsten}), 
\bea
\sten		&=&	s\nten+\Rten/T
\label{eqVsten}\;.
\eea
Following the phenomenological Israel-Stewart approach, $s=s_0$\/ remains related to $T=T_0$\/ 
by the Gibbs equation (\ref{eqGibbs}).
The dissipative part $\Rten$\/ is assumed to be an algebraic function of $\nten$\/ and
$\Tten$\/ only, which vanishes in equilibrium ($\Rten_0=0$\/). 
The theories of standard irreversible thermodynamics and of extended causal thermodynamics 
differ in the forms of $\Rten$\/, as given below.

The equations of energy and momentum conservation for (\ref{eqVTten}) can be written as
\bea
0		&=&	\Ddot\rho
			+\left(\rho+p+\mPi\right)\mTheta
			+\sigmaten:\piten
			+(2\aten+\Dten)\cdot\qten
\label{eqVEnergy}\;,\\
0		&=&	\left(\rho+p+\mPi\right)\aten
			+\Dten(p+\mPi)
			+\Dten\cdot\piten
			+\piten\cdot\aten
			+(\Ddot\qten)_\hten
			+\left(
			\sigmaten
			+\omegaten
			+4/3\mTheta\hten
			\right)\cdot\qten
\label{eqVMomentum}
\eea
where $\Bten:\Cten\equiv\tr(\Bten\cdot\Cten)$\/ for 2-tensor fields $\Bten$\/, $\Cten$\/.
The above treatment applies for a single-component fluid and allows a natural extension 
to multi-component fluids.
\subsection{Standard thermodynamics}
\label{sec2_4}
Standard thermodynamics assumes a linear dependence of $\Rten$\/ on the {\it thermodynamic fluxes} 
$\{\mPi,\qten,\piten\}$\/, which is possible only if (\ref{eqVsten}) takes the form
\bea
\sten		&=&	s\nten+\qten/T
\label{eqSIsten}\;.
\eea
Using (\ref{eqntenCons}), (\ref{eqVEnergy}) and (\ref{eqVMomentum}) yields the entropy generation rate
\bea
T\nabla\cdot\sten	
		&=&	-\mTheta\mPi
			-\left(\Dten\ln T+\aten\right)\cdot\qten
			-\sigmaten:\piten
\label{eqSIsGen}\;.
\eea
%In the limiting case of vanishing thermodynamic fluxes (\ref{eq11}) reduces to (\ref{eq6}).
The simplest relation to be imposed between the thermodynamic fluxes 
$\{\mPi,\qten,\piten\}$\/ and the {\it thermodynamic forces} 
$\{3\mTheta,\Dten\ln T+\aten,\sigmaten\}$\/ in agreement with (\ref{eqVsLaw}) 
is linear,
\bea
\mPi		&=&	-\zeta\mTheta
\label{eqSIPi}\;,\\
\qten		&=&	-\lambda T\left(\Dten\ln T+\aten\right)
\label{eqSIq}\;,\\
\piten		&=&	-2\eta\sigmaten
\label{eqSIpi}\;,
\eea
with non-negative coefficients of bulk viscosity $\zeta(\rho,n)$\/, thermal conductivity,
$\lambda(\rho,n)$\/ and shear viscosity $\eta(\rho,n)$\/.
This brings (\ref{eqSIsGen}) to
\bea
T\nabla\cdot\sten	
		&=&	\mPi^2/\zeta
			+\qten\cdot\qten/(\lambda T)
			+\piten:\piten/(2\eta)
\label{eqSIsGen2}\;,
\eea
which on using (\ref{eqntenCons}), (\ref{eqGibbs}), (\ref{eqVEnergy}) and (\ref{eqVMomentum}) 
yields an evolution equation for the entropy density $s$\/,
\bea
Tn\Ddot s	&=&	-\mTheta\mPi
			-\nabla\cdot\qten
			-\aten\cdot\qten
			-\piten:\sigmaten
\label{eqSIsEvol}\;.
\eea	
As a consequence of (\ref{eqSIPi})-(\ref{eqSIpi}), the thermodynamic fluxes
react instantaneously to the corresponding thermodynamic forces,
implying propagation of signals at (causality violating) infinite speed.
In standard thermodynamics thedynamics of the fluid is governed by the
(compressible) Navier-Stokes equation, which results from substitution of
(\ref{eqSIPi})-(\ref{eqSIpi}) into (\ref{eqVMomentum}).
\subsection{Causal thermodynamics}
\label{sec2_5}
Kinetic theory can motivate that $\Rten$\/ is second-order in the dissipative terms
\cite{IsraelStewart79}.
Truncation at first order removes terms necessary for causality and stability.
The most general algebraic form for $\Rten$\/ of at most second-order in the 
dissipative fluxes leads to 
\bea
\sten		&=&	s\nten
			+\qten/T
			-1/2\left(
			\beta_0\mPi^2
			+\beta_1\qten\cdot\qten
			+\beta_2\piten:\piten\right)\uten/T
			+\alpha_0\mPi\qten/T
			+\alpha_1\piten\cdot\qten/T
\label{eqECsten}\;.
\eea
The entropy density measured in LRF then becomes
\bea
-\sten\cdot\uten
		&=&	sn
			-1/2\left(
			\beta_0\mPi^2
			+\beta_1\qten\cdot\qten
			+\beta_2\piten:\piten\right)
\label{eqECsPhys}\;.
\eea
The negative sign of the non-equilibrium contributions reflects the fact that
the entropy density is maximum in equilibrium. 
The thermodynamic coefficients $\beta_j(\rho,n)\ge0$\/ in (\ref{eqECsten}) model 
deviations of the physical entropy density from $sn$\/ due to scalar/vector/tensor dissipative 
contributions to $\Rten$\/. 
The $\alpha_i(\rho,n)$\/ model contributions due to viscous/heat coupling,
which do not influence the physical entropy density (\ref{eqECsPhys}). 

The entropy generation rate associated with (\ref{eqECsten}) follows from 
(\ref{eqntenCons}), (\ref{eqGibbs}),  (\ref{eqVEnergy}) and (\ref{eqVMomentum}) to
\bea
T\nabla\cdot\sten
		&=&	-\mPi X
			-\qten\cdot\Yten
			-\piten:\langle\Zten\rangles
\label{eqECsGen}
\eea
with the scalar, vector and rank-2 tensor fields 
\bea
X		&=&	\mTheta
			+\beta_0\Ddot\mPi
			-\alpha_0\Dten\cdot\qten
			-\kappa_0T\qten\cdot\nabla\left(\alpha_0/T\right)
			+1/2\mPi T\nabla\cdot\left(\beta_0\uten/T\right)
\label{eqECX}\;,\\
\Yten		&=&	\nabla\ln T
			+\aten
			+\beta_1\Ddot\qten
			-\alpha_0\nabla\mPi
			-\alpha_1\nabla\cdot\piten\nonumber\\
		&&	-(1-\kappa_0)\mPi T\nabla\left(\alpha_0/T\right)
			-(1-\kappa_1)T\piten\cdot\nabla\left(\alpha_1/T\right)
			+1/2T\qten\nabla\cdot\left(\beta_1\uten/T\right)
\label{eqECY}\;,\\
\Zten		&=&	\nabla\uten
			+\beta_2\Ddot\piten
			-\alpha_1\nabla\qten
			-\kappa_1T\qten\otimes\nabla\left(\alpha_1/T\right)
			+1/2T\piten\nabla\cdot\left(\beta_2\uten/T\right)
\label{eqECZ}\;.
\eea	
The simplest {\it evolution equations} for the {\it causal thermodynamic fluxes} $\{\mPi,\qten,\piten\}$\/ 
in agreement with the second law (\ref{eqVsLaw}) are again linear relationships,  
\bea
\mPi		&=&	-\zeta X
\label{eqECPi}\;,\\
\qten		&=&	-\lambda T\Yten_\uperp
\label{eqECq}\;,\\
\piten		&=&	-2\eta\langle\Zten\rangle^{\rm s}_\uperp
\label{eqECpi}\;.
\eea
Two additional thermodynamic coefficients $\kappa_k(\rho,n)$\/ had to be introduced
in (\ref{eqECX})-(\ref{eqECZ})
as a consequence of the ambiguity involved in factoring terms which involve products 
$\mPi\qten$\/ and $\piten\cdot\qten$\/ in (\ref{eqECsten}).
Furthermore, (\ref{eqECPi})-(\ref{eqECpi}) contain terms involving gradients of the 
$\alpha_i$\/ and $\beta_j$\/.
Since the $\kappa_k$\/ are unknown a priori, these terms could be important even 
if the gradients themselves are small \cite{HiscockLindblom83}.
Finally, in (\ref{eqECX})-(\ref{eqECZ}) we neglected 
further contributions due to additional coupling terms between $\{\Pi,\qten,\piten\}$\/ and 
$\aten$\/, $\omegaten$\/, which can be shown to exist in kinetic theory \cite{IsraelStewart79}.

The complexity of the full evolution equations (\ref{eqECPi})-(\ref{eqECpi}) makes 
applications tractable only if certain simplifications are made.
%This requires both the corresponding thermodynamic scalar and the corresponding gradient of 
%$\alpha_i$\/ or $\beta_j$\/ to be small quantities.
A particularly simple set of evolution equations results from the assumptions \cite{Maartens97} 
\bea
0               &=&     \kappa_0
                =       \kappa_1       
\label{eqTECkappa}\;,\\
0               &=&     \alpha_0
                =       \alpha_1
\label{eqTECalpha}\;,\\
0               &\simeq&\nabla\cdot(\beta_j\uten/T)
\label{eqTECbeta}\;,
\eea
where (\ref{eqTECkappa}) reflects essentially the lack of knowledge on
the $\kappa_k$\/ while (\ref{eqTECalpha}) neglects the coupling between heat flux
and viscosity. Implications of (\ref{eqTECbeta}) are multifold and need to be justified
after a particular solution was found using a parametrization for $\beta_j(\rho,n)$\/.
The evolution equations resulting from (\ref{eqECPi})-(\ref{eqECpi}) under the assumptions 
(\ref{eqTECkappa})-(\ref{eqTECbeta}) are of covariant relativistic Maxwell-Cattaneo form, 
\bea
\tau_0(\Ddot\mPi)_\uperp+\mPi		&=&	\SIPi
\label{eqTECPi}\;,\\
\tau_1(\Ddot\qten)_\uperp+\qten		&=&	\SIqten	
\label{eqTECq}\;,\\
\tau_2(\Ddot\piten)_\uperp+\piten	&=&	\SIpiten
\label{eqTECpi}\;,
\eea
with the relaxation times $\tau_j(\rho,n)$\/ given by
\beq
\tau_0		=	\zeta\beta_0\sks
\tau_1		=	\lambda T\beta_1\sks
\tau_2		=	2\eta\beta_2\;,
\label{eqTauDef}\;,
\eeq
and $\{\SIPi,\SIqten,\SIpiten\}$\/ the (re-named) standard thermodynamic fluxes as in
(\ref{eqSIPi})-(\ref{eqSIpi}).
In contrast to the algebraic constraint equations (\ref{eqSIPi})-(\ref{eqSIpi}),
the evolution equations (\ref{eqTECPi})-(\ref{eqTECpi}) are first order partial 
differential equations,
which assure that in the LRF the viscous bulk/shear stresses and the heat flux relax
towards their standard limits $\{\SIPi,\SIqten,\SIpiten\}$\/ on time-scales $\tau_j$\/.
The relaxation times $\tau_j$\/ follow in principle from kinetic theory,
but can be estimated as mean collision times, $1/\tau\sim n\mSigma v$\/, 
with $\mSigma$\/ the collision cross section and $v$\/ the mean particle speed.
For later use we re-write (\ref{eqTECPi})-(\ref{eqTECpi}) as
\bea
\Ddot\mPi               &=&     \fractau{0}\Big(\SIPi-\mPi\Big)
\label{eqTECPiNew}\;,\\
\Ddot\qten              &=&     \fractau{1}\Big(\SIqten-\qten\Big)
                                +(\qten\cdot\aten)\uten
\label{eqTECqNew}\;,\\
\Ddot\piten             &=&     \fractau{2}\Big(\SIpiten-\piten\Big)
                                +\uten\otimes(\piten\cdot\aten)
                                +(\piten\cdot\aten)\otimes\uten
\label{eqTECpiNew}\;.
\eea
 
The conservation laws (\ref{eqntenCons}), (\ref{eqVEnergy}), (\ref{eqVMomentum})
and the Einstein equations (\ref{eqEinstein}), together with
the evolution equations (\ref{eqECPi})-(\ref{eqECpi}) constitute a complete system 
of hyperbolic first order PDEs for the solution vector of 24 (=1+1+3+10+1+3+5) dynamical variables 
$\{n$\/, $\rho$\/, $\uten$\/, $\gten$\/, $\mPi$\/, $\qten$\/, $\piten\}$\/.
This system represents a causal and stable theory for dissipative fluids
\cite{HiscockLindblom83}.
\subsection{Weakly dissipative fluids}
\label{sec2_6}
In many applications the inertia due to the dissipative contributions 
$\{\mPi,\qten,\piten\}$\/ can be neglected.
In addition, it is convenient to simplify the evolution equations (\ref{eqTECPi})-(\ref{eqTECpi}),
which depend on the kinematic properties $\{\mTheta,\aten,\sigmaten\}$\/ of the fluid,
among which $\sigmaten$\/ is particularly expensive to compute. 
An appropriate simplification can be obtained by calculating (\ref{eqTECPi})-(\ref{eqTECpi})
under the assumption of vanishing acceleration. 
Thermodynamic fluxes and thermodynamic forces calculated in this limit are underlined
in the following.
Note that the assumption of geodesic trajectories is only made for the calculation of the
dissipative terms and not for the dynamics, i.e. we do not assume a geodesic velocity 
field satisfying $\tnabla\uten\uten=0$\/, but rather leave $\uten$\/ unspecified.
In the following this will be referred to as the weakly dissipative limit.
In this limit the standard constraint equations (\ref{eqSIPi})-(\ref{eqSIpi}) reduce to
\bea
\GSIPi		&=&	-\zeta\GTheta
\label{eqGSIPi}\;,\\
\GSIqten	&=&	-\lambda T\Dten\ln T
\label{eqGSIqten}\;,\\
\GSIpiten	&=&	-2\eta\Gsigmaten
\label{eqGSIpiten}\;,
\eea
with the thermodynamic forces calculated from the kinematic properties
\beq
\GTheta         \equiv       	\mTheta|_{\aten=0}\sks
\Gaten          \equiv       	\aten|_{\aten=0}=0\sks
\Gsigmaten      \equiv       	\sigmaten|_{\aten=0}
		=		\langle\nabla\uten\rangles-1/3\GTheta\hten
\label{eqGkin}\;.
\eeq
The causal evolution equations (\ref{eqTECPiNew})-(\ref{eqTECpiNew}) simplify to
\bea
\Ddot\GPi 	&=&	\fractau{0}\Big(\GSIPi-\GPi\Big)
\label{eqGTECPi}\;,\\
\Ddot\Gqten	&=&	\fractau{1}\Big(\GSIqten-\Gqten\Big)
\label{eqGTECq}\;,\\
\Ddot\Gpiten	&=&	\fractau{2}\Big(\GSIpiten-\Gpiten\Big)
\label{eqGTECpi}\;.
\eea
\section{Dissipative hydrodynamics in 3+1 formulation}
\label{sec3}
Particularly useful for time-dependent calculations in general relativity is the 3+1 
formulation, where time derivatives are always with respect to globally defined universal time. 
Applications of 3+1 hydrodynamics and 3+1 magnetohydrodynamics have been mostly
restricted to ideal fluids (see Bonazzola et al. 1993 for an exception).
A collection of numerous general relativistic equations in the 3+1 representation
can be found in Durrer \& Straumann \shortcite{DurrerStraumann88}.
We give a 3+1 representation of relativistic dissipative hydrodynamics for
both, standard and causal thermodynamics.
For a detailed derivation of the equations presented in this section we refer to
Peitz \cite{Peitz98}.
\subsection{Generalities on the 3+1 formalism}
\label{sec3_1}
Assuming that spacetime \ST admits a slicing by slices $\Sigma_t$\/,
i.e. there is a diffeomorphism $\Phi:{\cal M}\mapsto\Sigma\times I,I\subset\bbbr$\/,
such that the manifolds $\Sigma_t=\Phi^{-1}(\Sigma\times\{t\})$\/ are spacelike
and the curves $\Phi^{-1}(m,t)$\/ are timelike. 
These curves define a vector field $\part$\/ which can be decomposed into normal 
and parallel components relative to the slicing,
\bea
\part      	&=&     \alpha\hnten+\tbeta
\label{eqSlicing}\;.
\eea
Here $\hnten$\/ is the timelike unit normal field (congruence of fiducial observers=FIDOs) 
and $\tbeta$\/ is tangent to the slices $\Sigma_t$\/.
$\alpha$\/ is the lapse function and $\tbeta$\/ is the shift vector field.
A coordinate system $\{x^i\}$\/ on $\Sigma$\/ induces natural coordinates on $\cal M$\/,
i.e. $\Phi^{-1}(m,t)$\/ has coordinates $(t,x^i)$\/ if $m\in\Sigma$\/ has coordinates $x^i$\/.
The timelike curves $\part$\/ have constant spatial coordinates (preferred timelike curves).
Now set $\tbeta=\beta^i\partial_i$\/ (where $\partial_i\equiv\partial/\partial x^i$\/). 
From $\gten(\hnten,\partial_i)=0$\/ one finds
$\gten(\partial_t,\partial_t)=-(\alpha^2-\beta^i\beta_i)$\/ and 
$\gten(\partial_t,\partial_i)=\beta_i$\/.
In coordinates co-moving with the FIDOs the metric thus reads
\bea
\gten           &=&     -(\alpha^2-\beta^i\beta_i)\dten t\otimes\dten t
                        +\beta_i\dten t\otimes\dten x^i
                        +\beta_i\dten x^i\otimes\dten t
                        +\gamma_{ij}\dten x^i\otimes\dten x^j
\label{eqgtenGen1}\\
           	&=&     -\alpha^2\dten t\otimes\dten t
                        +\gamma_{ij}(\dten x^i+\beta^i\dten t)\otimes(\dten x^j+\beta^j\dten t)
\label{eqgtenGen2}\;.
\eea
The forms $\dten t$\/ and $\dten x^i+\beta^i\dten t$\/ are thus orthogonal.
$\tgamma$\/ is the metric induced on $\Sigma_t$\/, and the affine connection on \AS is
denoted by $\tnabla$\/.

The tangent and cotangent spaces of $\cal M$\/ have two natural decompositions,
which give rise to two types of bases of vector fields and 1-forms. 
These are the dual pair $\{\partial_\mu\}$\/ and $\{\dten x^\mu\}$\/ for comoving 
coordinates $\{x^\mu\}$\/ and, on the other hand, the dual pair
$\{\hnten,\partial_i\}$\/ and $\{\alpha\dten t,\dten x^i+\beta^i\dten t\}$\/.
Instead of the coordinate basis $\{\partial_i\}$\/ one may also use an orthonormal 
horizontal basis $\{\te_i\}$\/ with $\gten(\te_i,\te_j)=\delta_{ij}$\/,
together with the dual basis $\{\ttheta^j\}$\/ for $\{\dten x^i\}$\/.
Then one has the two dual pairs
$\{\partial_t,\te_i\}$\/, $\{\dten t,\ttheta^i\}$\/ and
$\{\eten_0=\hnten,\te_i\}$\/, $\{\thetaten^\mu\}$\/ with
the orthonormal tetrad $\{\thetaten^\mu\}$\/ given by
$\thetaten^0=\alpha\dten t$\/ and $\thetaten^i=\ttheta^i+\beta^i\dten t$\/,
with $\beta^i$\/ defined by $\tbeta=\beta^i\te_i$\/. 
From (\ref{eqSlicing}) follows the relation
\beq
\eten_0         =     \hnten     
                =     \fraca\left(\partial_t-\tbeta\right)
\label{eqeten0}\;.
\eeq
The 3+1 representation of respectively a scalar field $A$\/, a vector field $\Bten$\/ 
and a symmetric rank-2 tensor field $\Cten$\/ defined on \ST is understood by
a representation with respect to the basis $\{\eten_0=\hnten,\te_i\}$\/, 
which we shall write in the following as
\bea
A		&=&	\AE
\label{eqScalarSplit}\;,\\
\Bten		&=&	\BE\eten_0+\BtS
\sks
\BtS=\BS^i\te_i
\label{eqVectorSplit}\;,\\
\Cten		&=&	\CE\eten_0\otimes\eten_0
			+\CtS\otimes\eten_0
			+\eten_0\otimes\CtS
			+\CtT
\sks
\CtS		=	\CS^i\te_i
\sks
\CtT		=	\CT^{ij}\te_i\otimes\te_j
\label{eqTensorSplit}\;.
\eea
The vector field $\BtS$\/ on \AS corresponding to the vector field $\Bten$\/
on \ST is referred to as the horizontal part of $\Bten$\/ and,
accordingly, the tensor field $\CtT$\/ on \AS is the horizontal part of $\Cten$\/.
Horizontal tensor fields appear bold face.
They can be viewed as spatial components of the fields on \ST
(which appear as a subscript), after having been projected onto the 3-space
orthogonal to $\hnten$\/ by $(\,)_{\gammaten}$\/.
If $\Bten=\BtS$\/, $\Bten$\/ is said to be a spatial vector field and, respectively,
$\Cten$\/ is a spatial tensor field if $\Cten=\CtT$\/.

The 3+1 representation of the affine connection $\nabla$\/ on \ST depends on the
connection forms, which depend on the kinematic properties of the FIDO congruence 
according to the irreducible decomposition of $\nabla\hnten$\/ in the sense of (\ref{eqDecomp}).
The FIDO's kinematic properties are distinguished from kinematic properties of
the fluid by a hat, i.e. $\hsigmaten$\/, $\homegaten$\/, $\haten$\/ and $\hat\mTheta$\/
are the shear, vorticity, acceleration and expansion of the FIDO congruence, respectively.
The fields $\hsigmaten$\/, $\homegaten$\/ and $\haten$\/ live in \AS and are therfore
spatial tensor fields, which we denote by $\thsigma\equiv\tTX_\hsigmaten$\/, 
$\thomega\equiv\tTX_\homegaten$\/ and $\tha\equiv\tSX_\haten$\/.
The induced metric $\tgamma$\/ on \AS is the horizontal part of the projector
$\gammaten\equiv\gten+\hnten\otimes\hnten$\/ into the FIDO's frame.

The FIDO world lines are orthogonal to the hypersurfaces and thus rotationsfree, 
$\homegaten=\thomega=0$\/.
The FIDO's acceleration $\tha$\/ is related to the lapse function by $\tha=\tnabla\ln\alpha$\/.
The connection forms on \ST can then be written in terms of the horizontal connection forms on
\AS, and the horizontal parts of only two spatial tensor fields, namely
\bea
\tha 		&=&      	\tnabla\ln\alpha
\label{eqGdef}\;,\\
\tK             &=&        	\tTX_\Kten
\label{eqKdef}\;.
\eea
Here $\Kten$\/ is the extrinsic curvature tensor (second fundamental form), 
defined on \ST by
\bea
\Kten		&\equiv&	-\nabla\hnten
		=		-\frac{1}{2}\lie_\hnten\gammaten
\label{eqKten}\;.
\eea
An equation for $\tK$\/ is obtained from the 3+1 representation of Einstein`s equation
(cf. (\ref{eq3+1EinsteinCurvature}) below).
The 3+1 representation of the second equality of (\ref{eqKten}) is also
recovered in the 3+1 formulation of Einstein's equations (cf. (\ref{eq3+1EinsteinMetric}) below),
and provides an equation for $\tgamma$\/.
This equation may be written in an alternative form based on the 3+1 representation
of the first equality in (\ref{eqKten}), namely as
\bea
\tK		&=&	-\Big(
			\thsigma
			+\frac{1}{3}\hTheta\tgamma
			+\frac{1}{2\alpha}\partial_t\tgamma\Big)
\label{eqtKalt}\;,
\eea
which confirms that $\Kten$\/ is indeed spatial (recall that $\thomega=0$\/) and,
furthermore, that $\tK$\/ is symmetric, $\tK=\langle\tK\rangles$\/.

For later use we give the 3+1 representation of $\Ddot A$\/, $\Ddot\Bten$\/ and
$\Ddot\Cten$\/ according to (\ref{eqScalarSplit})-(\ref{eqTensorSplit}),
\bea
\EX_{\Ddot A}           &=&     \gDConLie\EX_A
                                +\Dbul\EX_A
\label{eqScalarDotE}\;,\\
\EX_{\Ddot\Bten}        &=&     \gDConLie\EX_\Bten
                                +\Dbul\EX_\Bten
                                -\tF\cdot\tSX_\Bten
\label{eqVectorDotE}\;,\\
\tSX_{\Ddot\Bten}       &=&     \gDConLie\tSX_\Bten
                                +\Dbul\tSX_\Bten
                                -\EX_\Bten\tF
                                -\gam\tK\cdot\tSX_\Bten
\label{eqVectorDottS}\;,\\
\EX_{\Ddot\Cten}        &=&     \gDConLie\EX_\Cten
                                +\Dbul\EX_\Cten
                                -2\tF\cdot\tSX_\Cten
\label{eqTensorDotE}\;,\\
\tSX_{\Ddot\Cten}       &=&     \gDConLie\tSX_\Cten
                                +\Dbul\tSX_\Cten
                                -\gam\tK\cdot\tTX_\Cten
                                -\left(\EX_\Cten\tgamma+\tTX_\Cten\right)\cdot\tF
\label{eqTensorDottS}\;,\\
\tTX_{\Ddot\Cten}       &=&     \gDConLie\tTX_\Cten
                                +\Dbul\tTX_\Cten
                                -2\gam\tK\cdot\tTX_\Cten
                                -\tF\otimes\tSX_\Cten
                                -\tSX_\Cten\otimes\tF
\label{eqTensorDottT}\;,
\eea
with the horizontal vector field $\tF$\/ defined on \AS by
\bea
\tF                     =       -\gam\tnabla\ln\alpha
                                +\tK\cdot\tgv
\label{eqtF}\;.
\eea
$\Liebeta$\/ is the Lie derivative on \AS with respect to $\tbeta$\/.
The horizontal projection operator $\ten{h}=\tgamma+\tgv\otimes\tgv$\/ 
defined on \AS projects into the 2-space orthogonal to $\tgv$\/.
Operators $(\;)_\vpara$\/ and $(\;)_\vperp$\/, 
$\Dbul\equiv\tnabla_\vpara=\tgv\cdot\tnabla$\/
and $\tD\equiv\tnabla_\vperp=\ten h\cdot\tnabla$\/
as well as $\bf\langle\;\bf\rangle^{\rm s}$\/ and $\bf\langle\;\bf\rangle^{\rm a}$\/
on \AS are to be understood in analogy to the corresponding operators on \ST.
According to the irreducible decomposition
\bea
\tnabla\tgv     &=&     \tsigma
                        +\tomega
                        +\frac{1}{3}\vartheta\ten h
                        -\ta\otimes\tgv
\label{eqDecompAS}\;,
\eea
the kinematic properties of the fluid on \AS are
given by $\tsigma\equiv\langle\tnabla\tgv\rangle^{\rm s}_\vperp$\/,
$\tomega\equiv\langle\tnabla\tgv\rangle^{\rm a}_\vperp$\/,
$\ta\equiv\Dbul\tgv$\/ and $\vartheta\equiv\tnabla\cdot\tgv$\/.
%Notice that in the three-dimensional decomposition considered here there appears
%a factor $1/3$\/, as in the four-dimensional case. A factor $1/2$\/, which might
%have been expected, appears only in 2D. The reason for the equal factors in 3D and 4D
%is the equivalence principle.
%
\subsection{Conservation laws and Einstein's equations}
\label{sec3_2}
The velocity $\uten$\/ has the 3+1 representation
\bea
\uten           &=&     \gam(\eten_0+\tv)
		=	\gam\eten_0+\tgv
\label{equSplit}\;,
\eea
where $\gam\equiv\uE=-\hnten\cdot\uten$\/ is the Lorentz factor with respect to the FIDOs
and $\tgv\equiv\utS=\gam\tv$\/ is the horizontal part of 
$\gammaten\cdot\uten=(\gten+\hnten\otimes\hnten)\cdot\uten
=\uten+(\hnten\cdot\uten)\hnten=\uten-\gam\hnten$\/.
Since $\{\te_i\}$\/ is orthonormal, the $v^i$\/ are {\it physical}
3-velocity components as measured by FIDOs.
Similarly, the particle current $\nten$\/ has the 3+1 representation
\bea
\nten		&=&	n\gam(\eten_0+\tv)
		=	n(\gam\eten_0+\tgv)
\label{eqnSplit}\;.
\eea
The stress energy tensor $\Tten$\/ in (\ref{eqVTten}) has the 3+1 representation
\bea
\TE             &=&     (\rho+p+\mPi)\gam^2
                        -(p+\mPi)
                        +2\gam\qE
                        +\piE
\label{eq3+1VTE}\;,\\
\TtS            &=&     (\rho+p+\mPi)\gam\tgv
                        +\gam\qtS
                        +\qE\tgv
                        +\pitS
\label{eq3+1VTtS}\;,\\
\TtT            &=&     (\rho+p+\mPi)\tgv\otimes\tgv
                        +(p+\mPi)\tgamma
                        +\tgv\otimes\qtS
                        +\qtS\otimes\tgv
                        +\pitT\\
                &=&     \rho\tgv\otimes\tgv
                        +(p+\mPi)\ten h
                        +\tgv\otimes\qtS
                        +\qtS\otimes\tgv
                        +\pitT
\label{eq3+1VTtT}\;.
\eea
Note that $\qten$\/ and $\piten$\/ are orthogonal to $\uten$\/ but not to $\hnten$\/, 
and consequently $\qten$\/ and $\piten$\/ are no spatial fields.

The 3+1 representation of particle number conservation (\ref{eqntenCons}),
formally given by $0=E_{\nabla\cdot\nten}$\/, can be written as
\bea
0		&=&	\DConLie(\gam n)
		 	+\frac{1}{\alpha}\tnabla\cdot(\alpha n\tgv)
			-\gam n\tr(\tK)
\label{eq3+1Continuity}\;.
\eea
The 3+1 representation of stress-energy conservation (\ref{eqTtenCons}) splits into 
the energy equation, $0=E_{\nabla\cdot\Tten}$\/, 
and the momentum equation, $0=\tS_{\nabla\cdot\Tten}$\/, which can be regarded
as the spatial part of $0=\gammaten\cdot(\nabla\cdot\Tten)$\/. 
Leaving $\Tten$\/ unspecified, these are
\bea
0		&=&	\DConLie\TE
			-\tr(\tK)\TE
			+2(\TtS\cdot\tnabla)\ln\alpha
			+\tnabla\cdot\TtS
			-\tr(\tK\cdot\TtT)
\label{eq3+1Energy}\;,\\
0		&=&	\DConLie\TtS
			+\TE\tnabla\ln\alpha
			-2\tK\cdot\TtS
			-\tr(\tK)\TtS
			+\frac{1}{\alpha}\tnabla\cdot(\alpha\TtT)
\label{eq3+1Momentum}\;.
\eea			

For completeness we give the 3+1 representation of the Einstein equations
(e.g. Durrer \& Straumann 1988).
They consist of evolution equations for $\tgamma$\/ and $\tK$\/,
obtained from the definition of the extrinsic curvature (\ref{eqKten}) 
and from the (space, space) components of (\ref{eqEinstein}).
These dynamical equations are first order differential equations, 
\bea
\DConLie\tgamma		&=&	-2\tK
\label{eq3+1EinsteinMetric}\;,\\
\DConLie\tK		&=&	{\ten Ri}(\tgamma)
				-2\tK^2
				+\tr(\tK)\tK
				-8\pi\TtT
				+4\pi\Big(\tr(\tK)-\TE\Big)\tgamma
				-{\ten He}(\ln\alpha)
\label{eq3+1EinsteinCurvature}\;,
\eea
where $\tK^2\equiv\tK\cdot\tK$\/, $\ten Ri(\;)$\/ is the Ricci tensor and 
$\ten He(\;)$\/ is the Hessian on \AS, respectively.
The (time, time) and (time, space) components of (\ref{eqEinstein}) yield the Hamiltonian and
momentum constraints,
\bea
R-\tr(\tK)^2-\tr(\tK^2) 	&=&	16\pi\TE
\label{eq3+1EinsteinHamilton}\;,\\
\tnabla\cdot\tK-\tnabla\tr(\tK) &=&	8\pi \TtS
\label{eq3+1EinsteinMomentum}\;,
\eea
where $R$\/ is the curvature scalar on \AS.
\subsection{The standard constraint equations}
\label{sec3_3}
The 3+1 representation of the constraint equations (\ref{eqSIPi})-(\ref{eqSIpi})
for the thermodynamic fluxes $\{\SIPi$\/, $\SIqten$\/, $\SIpiten\}$\/ depends on the
3+1 representation of the kinematic properties $\mTheta,\aten,\sigmaten$\/ of the fluid,
which can be calculated to
\bea
\ThetaE		&=&    	\Whelp
			+\vartheta
                        -\gam\tr(\tK)
\label{eqThetaE}\;,\\
\aE                  
%		&=&	\gDConLie\gam
%			+(\tgv\cdot\tnabla)\gam
%			-\gam\tG\cdot\tgv
%                        -\tgv\cdot\tK\cdot\tgv
%\nonumber\\
              	&=&    	\gam\Whelp
			+\Dbul\gam
                        -K_\vpara
\label{eqaE}\;,\\
\atS            
%		&=&     \gDConLie\tgv
%			+(\tgv\cdot\tnabla)\tgv
%			-\gam^2\tG
%                        -2\gam\tK\cdot\tgv
%\nonumber\\
              	&=&    	\gam\tWhelp
			+\ta
\label{eqatS}\;,\\
\sigmaE        	&=&	\frac{1}{3}\left(\gam^2-1\right)\Big(2\Whelp-\vartheta+\gam\tr(\tK)\Big)
			+\gam\Big(\Dbul\gam-K_\vpara\Big)
\label{eqsigmaE}\;,\\
\sigmatS       	&=&	\frac{1}{2}(\gam^2-1)\tWhelp
			+\frac{\gam}{6}\Big(\Whelp-2\vartheta+2\gam\tr(\tK)\Big)\tgv
			+\frac{1}{2}\Big(\tD\gam+\gam\ta\Big)
			-\Big(\frac{1}{2}K_\vpara\tgamma+\tK\Big)\cdot\tgv
\nonumber\\	&=&	\frac{1}{2}(\gam^2-1)\tWhelp
			+\frac{1}{2}\Big(\Whelp\tgv+\tD\gam+\gam\ta\Big)
			-\frac{1}{3}\Big(\Whelp+\vartheta-\gam\tr(\tK)\Big)\tgv
			-\Big(\frac{1}{2}K_\vpara\tgamma+\tK\Big)\cdot\tgv
\label{eqsigmatS}\;,\\
\sigmatT        &=&     \frac{\gam}{2}\Big(\tWhelp\otimes\tgv+\tgv\otimes\tWhelp\Big)
			-\frac{1}{3}\Big(\Whelp-\gam\tr(\tK)\Big)\ten h
			+\tsigma
                        -\gam\tK
%   		&=&     +\gam/2\left[\tgv\otimes\tWhelp+\tWhelp\otimes\tgv\right]
%                        -1/3\left(\Whelp+\gam\hTheta\right)\ten h
\label{eqsigmatT}\;,
\eea
with time derivatives contained in 
%the scalar help function $\Whelp$\/ and the vector help function $\tWhelp$\/ on \AS,
\bea
\Whelp  	&\equiv&\DConLie\gam+(\tgv\cdot\tnabla)\ln\alpha
%                        +(\tgv\cdot\tnabla)\ln\alpha
\label{eqthetahelp}\;,\\
\tWhelp         &\equiv&\DConLie\tgv
			-2\tK\cdot\tgv
			+\gam\tnabla\ln\alpha
%                        +\gam\tnabla\ln\alpha
%			-2\tK\cdot\tgv
\label{eqtWhelphelp}\;.
\eea

Decomposing $\Dten\ln T$\/ finally yields the 3+1 representation of
the constraint equations (\ref{eqSIPi})-(\ref{eqSIpi}),
\bea
\SIPiE		&=&	-\zeta\ThetaE
\label{eqSIPiE}\;,\\
\SIqE		&=&	-\lambda T\left[
			\left(\gam^2-1\right)\DConLie\ln T
			+\gam\Dbul\ln T
			+\aE
			\right]
\label{eqSIqE}\;,\\
\SIqtS		&=&	-\lambda T\left[
			\gam\tgv\DConLie\ln T
			+\tD\ln T
			+\atS
			\right]
\label{eqSIqtS}\;,\\
\SIpiE		&=&	-2\eta\sigmaE
\label{eqSIpiE}\;,\\
\SIpitS		&=&	-2\eta\sigmatS
\label{eqSIpitS}\;,\\
\SIpitT		&=&	-2\eta\sigmatT
\label{eqSIpitT}\;.
\eea
\subsection{The causal evolution equations}
\label{sec3_4}
For the sake of simplicity the following discussion will be restricted 
to the causal evolution equations (\ref{eqTECPiNew})-(\ref{eqTECpiNew}).
Generalization to the full evolution equations (\ref{eqECPi})-(\ref{eqECpi}) is straightforward.
The required 3+1 representations of $\Ddot\mPi$\/, $\Ddot\qten$\/, $\Ddot\piten$\/ follow readily
from (\ref{eqScalarDotE})-(\ref{eqTensorDottT}).
%\bea
%\PidotE         &=&     \frac{\gam}{\alpha}\left(\part-\Liebeta\right)\PiE
%                        +\Dbul\PiE
%\label{eqPidotE}\;,\\
%\qdotE          &=&     \frac{\gam}{\alpha}\left(\part-\Liebeta\right)\qE
%                        +\Dbul\qE
%                        -\tF\cdot\qtS
%\label{eqqdotE}\;,\\
%\qdottS         &=&     \frac{\gam}{\alpha}\left(\part-\Liebeta\right)\qtS
%                        +\Dbul\qtS
%                        -\qE\tF
%                        -\gam\tK\cdot\qtS
%\label{eqqdottS}\;,\\
%\pidotE         &=&     \frac{\gam}{\alpha}\left(\part-\Liebeta\right)\piE
%                        +\Dbul\piE
%                        -2\tF\cdot\pitS
%\label{eqpidotE}\;,\\
%\pidottS        &=&     \frac{\gam}{\alpha}\left(\part-\Liebeta\right)\pitS
%                        +\Dbul\pitS
%                        -\piE\tF
%                        -\tK\cdot(\gam\pitS+\pitT\cdot\tgv)
%\label{eqpidottS}\;,\\
%\pidottT        &=&     \frac{\gam}{\alpha}\left(\part-\Liebeta\right)\pitT
%                        +\Dbul\pitT
%                        -2\gam\tK\cdot\pitT
%                        -\tF\otimes\pitS
%                        -\pitS\otimes\tF
%\label{eqpidottT}
%\eea
%with the vector help function $\tF$\/ on \AS defined by
%\bea
%\tF             &\equiv&        -\gam\tnabla\ln\alpha
%                                +\tK\cdot\tgv
%\label{eqtFDef}\;.
%\eea
Remains to decompose products $\qten\cdot\aten$\/ in (\ref{eqTECqNew}) and 
$\piten\cdot\aten$\/ in (\ref{eqTECpiNew}),
\bea
\qtaE		&=&	\qE\aE+\qtS\cdot\atS
\label{eqqtaE}\;,\\
\pitaE		&=&	\piE\aE+\pitS\cdot\atS
\label{eqpitaE}\;,\\
\pitatS		&=&	\aE\pitS+\pitT\cdot\atS
\label{eqpitatS}\;.
\eea	
The 3+1 evolution equations for the causal thermodynamic fluxes 
$\{\PiE$\/, $\qE$\/, $\qtS$\/, $\piE$\/, $\pitS$\/, $\pitT\}$\/ then follow 
from (\ref{eqTECPiNew})-(\ref{eqTECpiNew}) to
\bea
\gDConLie\PiE+\Dbul\PiE 	&=&	\fractau{0}\Big(\SIPiE-\PiE\Big)
\label{eqTECPiE}\;,\\
\gDConLie\qE+\Dbul\qE 		&=&	\fractau{1}\Big(\SIqE-\qE\Big)
%					+\gam\aE\qE
%					+(\gam\atS+\tF)\cdot\qtS
					+\gam\qtaE
					+\tF\cdot\qtS
\label{eqTECqE}\;,\\			
\gDConLie\qtS+\Dbul\qtS 	&=&	\fractau{1}\Big(\SIqtS-\qtS\Big)
%					+\qE(\aE\tgv+\tF)
%					+(\qtS\cdot\atS)\tgv
%					+\gam\tK\cdot\qtS
					+\qtaE\tgv
					+\qE\tF
					+\gam\tK\cdot\qtS
\label{eqTECqtS}\;,\\
\gDConLie\piE+\Dbul\piE 	&=&	\fractau{2}\Big(\SIpiE-\piE\Big)
%					+2\gam(\piE\aE+\pitS\cdot\atS)
%					+2\gam\tF\cdot\pitS
					+2\gam\pitaE
					+2\tF\cdot\pitS
\label{eqTECpiE}\;,\\
\gDConLie\pitS+\Dbul\pitS 	&=&	\fractau{2}\Big(\SIpitS-\pitS\Big)
%					+\gam(\aE\pitS+\pitT\cdot\atS)
%					+(\piE\aE+\pitS\cdot\atS)\tgv
					+\gam\pitatS
					+\pitaE\tgv
					+\gam\tK\cdot\pitS
					+\Big(\piE\tgamma+\pitT\Big)\cdot\tF
\label{eqTECpitS}\;,\\
\gDConLie\pitT+\Dbul\pitT 	&=&	\fractau{2}\Big(\SIpitT-\pitT\Big)
					+\tgv\otimes\pitatS
					+\pitatS\otimes\tgv
					+2\gam\tK\cdot\pitT
					+\tF\otimes\pitS
					+\pitS\otimes\tF
%				&&	+\tgv\otimes(\aE\pitS+\pitT\cdot\atS)
%					+(\aE\pitS+\pitT\cdot\atS)\otimes\tgv
\label{eqTECpitT}\;,
\eea			
with $\tF$\/ according to (\ref{eqtF}).
\subsection{The weakly dissipative limit}
\label{sec3_5}
The limit of weak dissipation discussed in section \ref{sec2_6} implies three
simplifications of the 3+1 equations of causal hydrodynamics.
The first simplfication concerns the 3+1 conservation laws (\ref{eq3+1Energy}) and 
(\ref{eq3+1Momentum}) for energy and momentum, where dissipative contributions 
$\{\mPi$\/, $\qten$\/, $\piten\}$\/ to the matter sources $\{\TE$\/, $\TtS$\/, $\TtT\}$\/ 
in (\ref{eq3+1VTE})-(\ref{eq3+1VTtT})
can be dropped wherever they enter algebraically, and only temporal and spatial gradients remain.
The second simplification affects the 3+1 representation of the kinematic
properties of the fluid, (\ref{eqThetaE})-(\ref{eqsigmatT}).
Using the geodesic conditions $\aE=\atS=0$\/, (\ref{eqaE}) and (\ref{eqatS})
yield simplified expressions for $\Whelp$\/ and $\tWhelp$\/, 
\bea
\Whelp			&=&	-\Dbul\ln\gam
				+K_\vpara/\gam
\label{eqWhelpGTEC}\;,\\
\tWhelp			&=&	-\ta/\gam
\label{eqtWhelpGTEC}\;,
\eea
which are no longer time-dependent. Therefore, any kinematic properties
(\ref{eqThetaE})-(\ref{eqsigmatT}) simplify to time-independent expressions
\bea
{_{\GTheta}E}		&=&	\vartheta
				-\Dbul\ln\gam
				+K_\vpara/\gam
				-\gam\tr(\tK)
\label{eqGThetaE}\;,\\
{_{\Gaten}E}		&=&	0
\label{eqGaE}\;,\\
{_{\Gaten}\tS}		&=&	0
\label{eqGatS}\;,\\
{_{\Gsigmaten}E}	&=&	\frac{1}{3}\left(\gam^2+2\right)
				\Big(\Dbul\ln\gam-K_\vpara/\gam\Big)
				-\frac{1}{3}\left(\gam^2-1\right)
				\Big(\vartheta-\gam\tr(\tK)\Big)
\nonumber\\		&=&	\frac{2}{3}\Big(\Dbul\ln\gam-K_\vpara/\gam\Big)
				+\frac{1}{3}\Big(\vartheta-\gam\tr(\tK)\Big)
				+\frac{1}{3}
				\Big(\Dbul\ln\gam-K_\vpara/\gam-\vartheta+\gam\tr(\tK)\Big)\gam^2
\label{eqGsigmaE}\;,\\
{_{\Gsigmaten}\tS}	&=&	\frac{1}{2}
				\Big(\ta/\gam+\tnabla\gam\Big)
				+\tK\cdot\tgv
				+\frac{1}{3}
				\Big(\Dbul\ln\gam-K_\vpara/\gam-\vartheta+\gam\tr(\tK)\Big)
				\gam\tgv	
%				(2-\gam^2)\ta/\gam
%				+\tnabla\gam
%				-2\tK\cdot\tgv
%				+2/3\left[\Dbul\ln\gam-K_\vpara/\gam-\vartheta-\gam\hTheta\right]
%				\gam\tgv 
\label{eqGsigmatS}\;,\\
{_{\Gsigmaten}\tT}	&=&	\underline{\tsigma}	
%				\tsigma
%				-\frac{1}{2}\left[\ta\otimes\tgv+\tgv\otimes\ta\right]
				-\gam\tK
				+\frac{1}{3}\Big(\Dbul\ln\gam-K_\vpara/\gam+\gam\tr(\tK)\Big)\ten h
\label{eqGsigmatT}\;,
\eea
with $\underline{\tsigma}=\langle\tnabla\tgv\rangles-1/2\vartheta\ten h$\/.
Finally, the weak dissipation limit affects the 3+1 evolution equations 
(\ref{eqTECPiE})-(\ref{eqTECpitT}) for the thermodynamic fluxes 
$\{\PiE$, $\qE$\/, $\qtS$\/, $\piE$\/, $\pitS$\/, $\pitT\}$\/, where products 
$\qtaE$\/, $\pitaE$\/, $\pitatS$\/ as in (\ref{eqqtaE})-(\ref{eqpitatS}) are dropped. 
This yields the weak dissipative evolution equations for the thermodynamic fluxes
$\{\GPiE$\/, $\GqE$\/, $\GqtS$\/, $\GpiE$\/, $\GpitS$\/, $\GpitT\}$\/,
\bea
\gDConLie\GPiE+\Dbul\GPiE 	&=&     \fractau{0}\Big(\GSIPiE-\GPiE\Big)
\label{eqGTECPiE}\;,\\
\gDConLie\GqE+\Dbul\GqE 	&=&     \fractau{1}\Big(\GSIqE-\GqE\Big)
                        		+\tF\cdot\GqtS
\label{eqGTECqE}\;,\\                    
\gDConLie\GqtS+\Dbul\GqtS 	&=&     \fractau{1}\Big(\GSIqtS-\GqtS\Big)
                        		+\GqE\tF
                        		+\gam\tK\cdot\GqtS
\label{eqGTECqtS}\;,\\
\gDConLie\GpiE+\Dbul\GpiE 	&=&     \fractau{2}\Big(\GSIpiE-\GpiE\Big)
                        		+2\tF\cdot\GpitS
\label{eqGTECpiE}\;,\\
\gDConLie\GpitS+\Dbul\GpitS 	&=&     \fractau{2}\Big(\GSIpitS-\GpitS\Big)
					+\gam\tK\cdot\GpitS
					+\Big(\GpiE\tgamma+\GpitT\Big)\cdot\tF
\label{eqGTECpitS}\;,\\
\gDConLie\GpitT+\Dbul\GpitT 	&=&     \fractau{2}\left[\GSIpitT-\GpitT\right]
                        		+2\gam\tK\cdot\GpitT
                        		+\tF\otimes\GpitS
                        		+\GpitS\otimes\tF
\label{eqGTECpitT}\;.
\eea                    
\section{Stationary and axisymmetric background spacetimes}
\label{sec4}
The 3+1 equations of dissipative hydrodynamics are specified
to the class of stationary, axisymmetric background spacetimes.
This situation is realized if the fluid under consideration has 
negligible influence on the gravitational field of a central object,
and in addition this field is known to be stationary and axisymmetric.
These assumptions include most applications related to accretion/ejection 
flows in the vicinity of compact objects.
For rotating black holes the vacuum metric is Kerr, 
and we give the equations also for this special case.
\subsection{Implications of symmetries}
\label{sec4_1}
The general form of a stationary, axisymmetric vacuum spacetime can be put in a form which is
symmetric under a simultaneous change of sign of $t$\/ and $\phi$\/, the Killing coordinates
associated with the commuting time and axial Killing vector fields $\kten$\/ and $\mten$\/.
Choosing the remaining two meridional coordinates as spherical coordinates allows to
write $\gten$\/ as
\bea
\gten		&=&	-\alpha^2\dten t\otimes\dten t
			+\tilomega^2\left(\dten\phi-\omega\dten t\right)
			\otimes\left(\dten\phi-\omega\dten t\right)
			+\exp{2\mu}\dten r\otimes\dten r
			+\exp{2\nu}\dten\theta\otimes\dten\theta
\label{eqgtenSym}\;,
\eea
with the invariant metric coefficients
\beq
\tilomega^2	=	\mten^2\sks
\omega		=	-\kten\cdot\mten/\mten^2\sks
\alpha^2	=	-\kten^2
			+\kten\cdot\mten/\mten^2\;.
\eeq
For the physical interpretation of $\tilomega$\/, $\omega$\/ and $\alpha$\/ see e.g. Bardeen 
\shortcite{Bardeen70}.
The generic choice of the fiducial congruence is (see e.g. Thorne \& Macdonald 
\shortcite{ThorneMacdonald82} for criteria that uniquely fix this choice)
\bea
\hnten          &=&     \eten_0
                =       \fraca\left(\kten+\omega\mten\right)
\label{eqhntenSym}\;.
\eea
These FIDOs possess vanishing specific angular momentum, ${\hat n}_\phi=\hnten\cdot\mten=0$\/. 
Therefore they correspond to Bardeen's \shortcite{Bardeen70} 
{\it zero angular momentum observers} (ZAMOs).
Furthermore, since $\hnten\cdot\mten=0$\/, $\mten$\/ is a spatial vector,
which we denote on \AS by $\tm\equiv\tSX_\mten$\/.
Comparison with (\ref{eqeten0}) shows that the shift vector has to be chosen as
\bea
\tbeta		&=&	-\omega\tm
\label{eqtbetaSym}\;,
\eea
and the metric induced on \AS becomes
\bea
\tgamma		&=&	\tilomega^2\dten\phi\otimes\dten\phi
			+\exp{2\mu}\dten r\otimes\dten r
			+\exp{2\nu}\dten\theta\otimes\dten\theta
\label{eqtgammaSym}\;.
\eea
The 3+1 representation of Killing's equation for $\kten$\/ and $\mten$\/, 
together with their commutivity, 
allows to establish the following relations \cite{ThorneMacdonald82}
\bea
\part\mu		&=&	0\sks
\part\nu		=	0\sks
\part\omega		=	0\sks
\part\tm		=	0\sks
\part\tgamma		=	0\;,\\
\tm\cdot\tnabla\mu	&=&	0\sks
\tm\cdot\tnabla\nu	=	0\sks
\tm\cdot\tnabla\omega	=	0\sks
\langle\tnabla\tm\rangles
			=	0\;,\\
\hTheta			&=&	0\sks
\thsigma		=	\frac{1}{2\alpha}\Big(
				\tm\otimes\tnabla\omega
				+\tnabla\omega\otimes\tm
				\Big)\;.
\eea
Note that $\langle\tnabla\tm\rangles=0$\/ states that $\tm$\/ is a Killing vector field on \AS.
However, as a consequence of $\tnabla\omega\neq0$\/, $\tbeta$\/ is not a Killing vector field on \AS.
As a concequence of $\partial_t\tgamma$\/ and $\hTheta=0$\/,
the horizontal field $\tK$\/ according to (\ref{eqtKalt}) reduces to
\bea
\tK		&=&	-\thsigma
		=	-\frac{1}{2\alpha}\Big(
			\tm\otimes\tnabla\omega
			+\tnabla\omega\otimes\tm\Big)
\label{eqtKSym}\;.
\eea
Therefore $\tK$\/ measures the shear of hypersurfaces $\Sigma_t$\/,
which vanishes for $\omega=0$\/.
\subsection{Gravitomagnetic and gravitoelectric tensor fields}
\label{sec4_2}
A characteristic phenomenon in axisymmetric spacetimes is the dragging of inertial frames.
Physical implications due to this effect (see Thorne et al. 1986 for the case of the 
Kerr metric) are often described in terms of the gravitomagnetic tensor field $\tH$\/, 
defined on \AS by
\bea
\tH		&\equiv&\fraca\tnabla\tbeta
		=	-\fraca\tnabla(\omega\tm)
		=	-\fraca\left[
			\omega\tnabla\tm
			+\tnabla\omega\otimes\tm
			\right]
\label{eqtH}\;.
\eea
It is clear from (\ref{eqtKSym}) that $\tK=\langle\tH\rangles$\/, and therefore
\bea
\tK\cdot\tT	&=&	\frac{1}{2}\left(\tH\cdot\tT+\tT\cdot\tH\right)
\label{eqtHuse}\;.
\eea
The antisymmetric part of $\tH$\/ can be expressed as an axial vector field on \AS, 
namely the gravitomagnetic vector field 
\bea
\tJ		&\equiv&	\fraca\tnabla\wedge\tbeta
\label{eqtJ}\;.
\eea
For an arbitrary vector field $\tS$\/ on \AS one finds the relation
\bea
\tH\cdot\tS	&=&	\tK\cdot\tS
			-\frac{1}{2}\tJ\wedge\tS
\label{eqtJuse}\;.
\eea
The Lie derivatives $\Liebeta$\/ in the 3+1 equations derived in section \ref{sec3}
can then be expressed as
\bea
\Liebeta E	&=&	(\tbeta\cdot\tnabla)E
\label{eqLiebetaE}\;,\\
\Liebeta\tS	&=&	(\tbeta\cdot\tnabla)\tS
			-(\tS\cdot\tnabla)\tbeta
		=	(\tbeta\cdot\tnabla)\tS
			-\alpha\tS\cdot\tH
\label{eqLiebetatS}\;,\\
\Liebeta\tT	&=&	(\tbeta\cdot\tnabla)\tT
			-2\langle(\tT\cdot\tnabla)\tbeta\rangles
		=	(\tbeta\cdot\tnabla)\tT
			-2\alpha\langle\tT\cdot\tH\rangles
		=	(\tbeta\cdot\tnabla)\tT 
			-2\tT\cdot\tK
\label{eqLiebetatT}\;.
\eea
The term $\tS\cdot\tH$\/ in (\ref{eqLiebetatS}) may be expressed by either
$\tH$\/ as in (\ref{eqtH}) or in terms of $\tJ$\/ and $\tK$\/ via (\ref{eqtJuse}).
In addition to the gravitomagnetic tensor fields $\tH$\/ and $\tJ$\/, 
we introduce the gravitoelectric vector field $\tG$\/, defined on \AS by
\bea
\tG		&\equiv&	-\tha
		=		-\tnabla\ln\alpha
\label{eqtG}\;.
\eea
This field measures the gravitational acceleration measured by the fiducial observers.
\subsection{Conservation laws}
\label{sec4_3}
The 3+1 conservation laws for particle number (\ref{eq3+1Continuity}), 
energy (\ref{eq3+1Energy}) and momentum (\ref{eq3+1Momentum}) in a
stationary, axisymmetric background read
\bea
0		&=&	\DConbeta(\gam n)
			+\fraca\tnabla(\alpha n\tgv)
\label{eq3+1ContinuitySym}\;,\\
0		&=&	\DConbeta\TE
			+\tnabla\cdot\TtS
			-2\TtS\cdot\tG
			-\tr(\tK\cdot\TtT)
\label{eq3+1TEnergySym}\;,\\
0		&=&	\DConbeta\TtS
			-\TE\tG
			-\tH\cdot\TtS
			+\fraca\tnabla(\alpha\TtT)\nonumber\\
		&=&	\DConbeta\TtS
			+\tnabla\cdot\TtT
			-\tH\cdot\TtS
			-\left(\TE\tgamma+\TtT\right)\cdot\tG	
\label{eq3+1MomentumSym}\;.
\eea
\subsection{Evolution equations}
\label{sec4_4}
The 3+1 representation (\ref{eqThetaE})-(\ref{eqsigmatT}) of the kinematic properties of
the fluid in stationary, axisymmetric background read
\bea
\ThetaE		&=&	\Whelp
			+\vartheta
\label{eqThetaESym}\;,\\
\aE		&=&	\gam\Whelp
			+\Dbul\gam
			+\hsigma_\vpara
\label{eqaESym}\;,\\
\atS		&=&	\gam\tWhelp
			+\ta
\label{eqatSSym}\;,\\
\sigmaE		&=&	\frac{3}{2}\left(\gam^2-1\right)\Big(\Whelp-2\vartheta\Big)
			+\gam\Big(\Dbul\gam+\hsigma_\vpara\Big)
\label{eqsigmaESym}\;,\\
\sigmatS	&=&	\frac{1}{2}\left(\gam^2-1\right)\tWhelp
			+\frac{\gam}{6}\Big(\Whelp-2\vartheta\Big)\tgv
			+\frac{1}{2}\Big(\tD\gam+\gam\ta\Big)
			-\Big(\hsigma_\vpara\tgamma+\thsigma\Big)\cdot\tgv
\label{eqsigmatSSym}\;,\\
\sigmatT	&=&	\frac{\gam}{2}\Big(\tWhelp\otimes\tgv+\tgv\otimes\tWhelp\Big)
			-\frac{1}{3}\Whelp\ten h
			+\tsigma
			+\gam\thsigma
\label{eqsigmatTSym}
\eea
with time derivatives contained in
\bea
\Whelp		&=&	\DConbeta\gam
			-G_\vpara
\label{eqWhelpSym}\;,\\
\tWhelp		&=&	\DConbeta\tgv
			-\tH\cdot\tgv
			-\gam\tG
\label{eqtWhelpSym}\;.
\eea
The 3+1 representation of the standard constraint equations (\ref{eqSIPiE})-(\ref{eqSIpitT}) 
can be written as
\bea
\SIPiE          &=&     -\zeta\ThetaE
\label{eqSIPiESym}\;,\\
\SIqE           &=&     -\kappa T\left[
                        \left(\gam^2-1\right)\DConbeta\ln T
                        +\gam\Dbul\ln T
                        +\aE
                        \right]
\label{eqSIqESym}\;,\\
\SIqtS          &=&     -\kappa T\left[
                        \gam\tgv\DConbeta\ln T
                        +\tD\ln T
                        +\atS
                        \right]
\label{eqSIqtSSym}\;,\\
\SIpiE          &=&     -2\eta\sigmaE
\label{eqSIpiESym}\;,\\
\SIpitS         &=&     -2\eta\sigmatS
\label{eqSIpitSSym}\;,\\
\SIpitT         &=&     -2\eta\sigmatT
\label{eqSIpitTSym}\;.
\eea
The 3+1 representation of the causal evolution equations (\ref{eqTECPiE})-(\ref{eqTECpitT}) reads
\bea
\gDConbeta\PiE+\Dbul\PiE 	&=&     \fractau{0}\Big(\SIPiE-\PiE\Big)
\label{eqTECPiESym}\;,\\
\gDConbeta\qE+\Dbul\qE 		&=&     \fractau{1}\Big(\SIqE-\qE\Big)
                        		+\gam\qtaE
                        		+\tF\cdot\qtS
\label{eqTECqESym}\;,\\
\gDConbeta\qtS+\Dbul\qtS 	&=&     \fractau{1}\Big(\SIqtS-\qtS\Big)
                        		+\qtaE\tgv
					+\qE\tF
                        		+\frac{1}{2}\gam\tJ\wedge\qtS
\label{eqTECqtSSym}\;,\\
\gDConbeta\piE+\Dbul\piE 	&=&     \fractau{2}\Big(\SIpiE-\piE\Big)
                        		+2\gam\pitaE
                        		+2\gam\tF\cdot\pitS
\label{eqTECpiESym}\;,\\
\gDConbeta\pitS+\Dbul\pitS	&=&     \fractau{2}\Big(\SIpitS-\pitS\Big)
                        		+\gam\pitatS
            	            		+\pitaE\tgv
					+\frac{1}{2}\gam\tJ\wedge\pitS
					+\left(\piE\tgamma+\pitT\right)\cdot\tF
\label{eqTECpitSSym}\;,\\
\gDConbeta\pitT+\Dbul\pitT 	&=&     \fractau{2}\Big(\SIpitT-\pitT\Big)
                		      	+\tgv\otimes\pitatS
                	        	+\pitatS\otimes\tgv
					+\gam\tF\otimes\pitS
					+\gam\pitS\otimes\tF
\label{eqTECpitTSym}\;,
\eea
with
\bea
\tF		&=&	\gam\tG-\thsigma\cdot\tgv
\label{eqtFSym}\;.
\eea
\subsection{Weakly dissipative limit}
\label{sec4_5}
The 3+1 representation of the kinematic properties of the fluid in the weak dissipation limit
can be written as 
\bea
{_{\GTheta}E}           &=&     \vartheta
                                -\Dbul\ln\gam
                                +\hsigma_\vpara/\gam
\label{eqGThetaESym}\;,\\
{_{\Gaten}E}            &=&     0
\label{eqGaESym}\;,\\
{_{\Gaten}\tS}          &=&     0
\label{eqGatSSym}\;,\\
{_{\Gsigmaten}E}        &=&     \frac{1}{3}\left(\gam^2+2\right)
                                \Big(\Dbul\ln\gam-\hsigma_\vpara/\gam\Big)
                                -\frac{1}{3}\left(\gam^2-1\right)\vartheta
\nonumber\\             &=&     \frac{2}{3}\Big(\Dbul\ln\gam-\hsigma_\vpara/\gam\Big)
                                +\frac{1}{3}\vartheta
                                +\frac{1}{3}
                                \Big(\Dbul\ln\gam-\hsigma_\vpara/\gam-\vartheta\Big)\gam^2
\label{eqGsigmaESym}\;,\\
{_{\Gsigmaten}\tS}      &=&     \frac{1}{2}
                                \Big(\ta/\gam+\tnabla\gam\Big)
                                +\thsigma\cdot\tgv
                                +\frac{1}{3}
                                \Big(\Dbul\ln\gam-\hsigma_\vpara/\gam-\vartheta\Big)\gam\tgv   
%                               (2-\gam^2)\ta/\gam
%                               +\tnabla\gam
%                               -2\tK\cdot\tgv
%                               +2/3\left[\Dbul\ln\gam-K_\vpara/\gam-\vartheta-\gam\hTheta\right]
%				\gam\tgv 
\label{eqGsigmatSSym}\;,\\
{_{\Gsigmaten}\tT}      &=&    	\underline{\tsigma} 
%				\tsigma
%                                -\frac{1}{2}\left[\ta\otimes\tgv+\tgv\otimes\ta\right]
                                -\gam\thsigma
                                +\frac{1}{3}\Big(\Dbul\ln\gam-\hsigma_\vpara/\gam\Big)\ten h
\label{eqGsigmatTSym}\;.
\eea
The 3+1 weak dissipation evolution equations (\ref{eqTECPiE})-(\ref{eqTECpitT}) for 
$\{\GPiE$\/, $\GqE$\/, $\GqtS$\/, $\GpiE$\/, $\GpitS$\/, $\GpitT\}$\/ can be written as
\bea
\gDConbeta\GPiE+\Dbul\GPiE       &=&     \fractau{0}\Big(\GSIPiE-\GPiE\Big)
\label{eqGTECPiESym}\;,\\
\gDConbeta\GqE+\Dbul\GqE         &=&     \fractau{1}\Big(\GSIqE-\GqE\Big)
                                        +\tF\cdot\GqtS
\label{eqGTECqESym}\;,\\                    
\gDConbeta\GqtS+\Dbul\GqtS       &=&     \fractau{1}\Big(\GSIqtS-\GqtS\Big)
                                        +\GqE\tF
                                        +\frac{1}{2}\gam\tJ\wedge\GqtS
\label{eqGTECqtSSym}\;,\\
\gDConbeta\GpiE+\Dbul\GpiE       &=&     \fractau{2}\Big(\GSIpiE-\GpiE\Big)
                                        +2\tF\cdot\GpitS
\label{eqGTECpiESym}\;,\\
\gDConbeta\GpitS+\Dbul\GpitS     &=&     \fractau{2}\Big(\GSIpitS-\GpitS\Big)
                                        +\frac{1}{2}\gam\tJ\wedge\GpitS
                                        +\Big(\GpiE\tgamma+\GpitT\Big)\cdot\tF
\label{eqGTECpitSSym}\;,\\
\gDConbeta\GpitT+\Dbul\GpitT     &=&     \fractau{2}\Big(\GSIpitT-\GpitT\Big)
                                        +\tF\otimes\GpitS
                                        +\GpitS\otimes\tF
\label{eqGTECpitTSym}
\eea                    
with $\tF$\/ according to (\ref{eqtFSym}).
\subsection{Specification to Kerr geometry}
\label{sec4_6}
The behaviour of fluids in the vicinity of a rotating black hole is governed by the conservation laws
and evolution equations in Kerr metric.
The Kerr metric is a two parameter family of stationary, axisymmetric vacuum spacetimes.
Parameters are the mass $M$\/ and the specific angular momentum $a\equiv J/M$\/ of the black hole. 
In terms of the functions
\bea
\mDelta			&\equiv&	r^2+a^2-2Mr\sks
\varrho^2		\equiv		r^2+a^2\cos^2\theta\sks
\mSigma^2		\equiv		(r^2+a^2)^2-a^2\Delta\sin^2\theta
\eea
the lapse function $\alpha$\/ and the non-vanishing components of the shift vector $\tbeta$\/
and the metric $\tgamma$\/ according to (\ref{eqgtenGen2}) are 
\bea
\alpha			&=&	\frac{\varrho\sqrt{\mDelta}}{\mSigma}\sks
\beta^\phi		=	-\omega
			=	-\frac{2aMr}{\mSigma^2}\sks
\tilomega		=	\frac{\mSigma}{\varrho}\sin\theta\sks
\exp{2\mu}		=	\frac{\varrho^2}{\mDelta}\sks
\exp{2\nu}		=	\varrho^2\;.
\eea
A consistent treatment of viscous hydrodynamics in the vicinity of a rotating black hole
would require a careful analysis of the boundary conditions at the horizon.
Such an analysis is best performed within the concept of a {\it streched horizon},
defined by a small value of lapse $\alpha$\/.
We refer to the {\it membrane paradigma} of Thorne et al. \shortcite{TPMSZ86},
where this approach is applied to ideal magnetohydrodynamics and Maxwell's equations.
This work also contains a coordinate representation of $\tG$\/ and $\tH$\/.
\subsection{Stationary and axisymmetric flows}
\label{sec4_7}
The assumptions of stationarity and axisymmetry so far concerned exclusively the back ground metric. 
In many problems it is justified to assume these symmetries to hold also for the fluid
configuration under consideration.
The equations for stationary flow are obtained by setting $\part=0$\/ in the 3+1 equations,
since now $\kten$\/ is a Killing vector field also for the fluid.
Similarly, the equations for axisymmetric flow follow from setting $\tbeta\cdot\tnabla=0$\/,
since $\tm$\/ then is a Killing vector field also or the fluid. 

The post-Newtonian limit of causal viscous hydrodynamics is recovered from the
3+1 equations by setting $\alpha=1$\/ ($\mapsto\tG=0$\/) and $\tbeta=0$\/ 
($\mapsto\tH=\thsigma=0$\/),
and furthermore setting $\gamma$\/ equal to unity, while gradients of $\gamma$\/ 
are retained.
This limit contains still non-vanishing quantities of post-Newtonian order $\nabla_\vpara\uten$\/.
Neglecting these then yields the Newtonian limit of causal viscous hydrodynamics,
where $\ThetaE$\/, $\aE$\/, $\sigmaE$\/, $\sigmatS$\/ vanish.
A related causal description of viscous angular momentum transfer in Newtonian accretion disc 
boundary layers was considered by Papaloizou \& Szuskiewicz 
\shortcite{PapaloizouSzuskiewicz94} and Kley \& Papaloizou \shortcite{KleyPapaloizou97}.
\section{Conclusions}
\label{concl}
We have provided a complete set of equations for dissipative 
relativistic hydrodynamics in their 3+1 representation.
Furthermore, we have specified the general system to the class of
stationary axisymmetric vacuum spacetimes, with the Kerr metric as the
most relevant astrophysical example.
For this case we have written the equations in a form where the dragging of
inertial frames is described by the gravitomagnetic tensor field.
This allows to combine the equations with the 3+1 Maxwell's equations
as in Thorne \& Macdonald \shortcite{ThorneMacdonald82}.

Causality has been accounted for by using the extended causal description 
of thermodynamics, relativistically formulated on a phenomenological level by 
Israel \shortcite{Israel76} and Stewart \shortcite{Stewart77} and justified 
by kinetic theory \cite{IsraelStewart79}.
In contrast to the conventionally used compressible Navier-Stokes description
of non-ideal hydrodynamics, the equations of extended causal thermodynamics guarantee
finite propagation speeds of heat and viscous signals and yield stable local 
thermodynamic equilibria \cite{HiscockLindblom83}. 

A causality preserving formulation is required whenever the
thermodynamic timescale becomes comparable to the dynamical timescale
and therefore the assumption of local thermodynamic equilibrium is
not justified.
This is particularly the case in supersonic flows and/or processes in the vicinity of 
the event horizon.
Astrophysical examples and thus potential fields for application of causal thermodynamics
in the formulation presented here include the gravitational collapse of stars 
(e.g. Baumgarte et al. 1995), 
the innermost parts of accretion discs around black holes 
(e.g. Peitz \& Appl 1997) or
the collision of neutron stars (e.g. Rasio \& Shapiro 1992).
In many problems of interest the inertia due to the dissipative
contributions to the stress-energy tensor can be neglected.
The 3+1 formulation of the corresponding simplified set of equations
is given.

The five conservation laws for particle number, energy and momentum,
the ten Einstein equations for the metric tensor and the ten evolution
equations for the thermodynamic fluxes form a hyperbolic system of first order PDEs
tractable by numerical methods (e.g. Bonazzola et al. \shortcite{Bonazzolaetal93}).

%%%%%%%%%%%%%%%%%%%%%%%%%%%%%%%%%%%%%%%%%%%%%%%%%%%%%%%%%%%%%%%%%%%%%%%%%%%%%%%%%%%%%%
%       acknowledgements
%%%%%%%%%%%%%%%%%%%%%%%%%%%%%%%%%%%%%%%%%%%%%%%%%%%%%%%%%%%%%%%%%%%%%%%%%%%%%%%%%%%%%%
%
\vspace{0.5cm}
We are grateful to M. Camenzind, R. Khanna and S. Spindeldreher for enlightening
discussions.
Support by Deutsche Forschungsgemeinschaft under SFB 328 (JP) and SFB 359 (SA),
and by the French Ministery of Foreign Affairs (SA) is acknowledged.

\end{document}